      [1999/12/01 v1.4c Il Nuovo Cimento]
\documentclass{cimento}

\usepackage{graphicx}

\newcommand{\GeV}{\,{\rm GeV}}

\title{QCD in hadron collisions\thanks{Contribution to the proceedings
    of the XXVI Rencontres de Physique de la Vall\'ee d'Aoste.}}
\author{Gavin~P.~Salam\from{ins:CERN}\from{ins:PU}\from{ins:Paris}
}
\instlist{%
  \inst{ins:CERN} CERN, PH-TH, CH-1211 Geneva 23, Switzerland.
  \inst{ins:PU}  Department of Physics, Princeton University, Princeton, NJ 08544, USA.
  \inst{ins:Paris} LPTHE; CNRS UMR 7589; UPMC Univ.\ Paris 6; Paris, France.
}
\PACSes{\PACSit{}{13.87.Ce,  13.87.Fh, 13.65.+i}}
\begin{document}

\maketitle

\begin{abstract}
  This talk examines recent progress in collider QCD and some facets
  of the interplay between these developments and searches for new
  particles and phenomena at the Tevatron and LHC.
\end{abstract}

\section{Introduction}

The past 18 months have seen considerable excitement in the field of
collider particle physics.
Among the topics that deserve a mention, one might include the
measurement of an unexpectedly large $t\bar t$ forward-backward
asymmetry at the Tevatron \cite{Aaltonen:2011kc,Abazov:2011rq}, CDF's
anomalous bump in W+dijet production~\cite{Aaltonen:2011mk}, the LHC's
exclusion of huge swathes of supersymmetric parameter
space~\cite{Aad:2011ib,Chatrchyan:2011zy} and hints of the Higgs boson from both the
Tevatron and the
LHC~\cite{ATLAS:2012ae,Chatrchyan:2012tx,TEVNPH:2012ab}.

The purpose of this talk is to examine a selection of the past years'
collider--QCD developments, attempting to place them in the context of
the above ``headline'' developments.
To guide us through this exercise, let us start by recalling that hard
collider events can be broken up into various components: the
non-perturbative structure of the proton, e.g.\ encoded in parton
distribution functions and as relevant also to multiple semi-soft
interactions; the ``hard'' process, amenable to fixed-order
perturbative calculation, usually involving at most a handful of
partons; the fragmentation of those partons, often implemented as a
parton shower; and the hadronisation process, by which partons turn
into the hadrons that are observed experimentally.
All of these elements must be dealt with correctly (and together) if
one is to fully predict the properties of events at hadron colliders.
We start our examination of recent QCD progress by considering the
element that has seen the most concerted effort, namely the hard
process.

\section{The hard process at next-to-leading order}

The hard process is where we make use of a perturbative expansion in
the strong coupling constant, $\alpha_s$.
Given that $\alpha_s \sim 0.1$ at the scales of relevance, one might
expect that a leading-order (LO) calculation should generally be
accurate to within about $10\%$.
It is quite straightforward to check whether this is the case across a
range of processes with the help of a tool such as
NLOJet++~\cite{Nagy:2003tz}, MCFM~\cite{Campbell:2002tg} and
VBFNLO~\cite{Arnold:2008rz}.
While there are cases where next-to-leading order corrections are
modest, for example for the inclusive jet spectrum, quite often one
finds that the NLO terms are large: the $Z$-boson $p_t$ spectrum sees
a $50\%$ correction at NLO; and the jet $p_t$ spectrum in events with
a $Z$-boson sees a NLO correction of several hundred percent.
Sometimes these large corrections can be understood based on simple
physical considerations, for example the appearance of new, enhanced
topologies at NLO.
There is, however, no general understanding of the different sizes of
the NLO corrections and so, to obtain a reasonable degree of accuracy
for hadron collider predictions, one must usually explicitly calculate
the NLO terms.

In recent years, guidance as to which NLO calculations to carry out
has come from the many
different searches that are being performed.
A classic example comes from SUSY searches: gluinos can decay to a
squark and antiquark, with the 
squark then decaying to a quark and unobserved neutralino,
so the signature for pair production of gluinos is the presence
of four jets and missing energy;
an important background is then $Z$ plus 4-jets, where the $Z$-boson
decays to neutrinos. One would like to know this background to NLO.

A measure of the difficulty of such a calculation is the number of
external legs: $Z$ plus 4-jet production is a $2\to5$ process.
The first hadron-collider NLO calculation was for the Drell-Yan
process in 1979, i.e.\ essentially a $2\to1$ process~\cite{DYNLO}.
It was not until 1987 that $2\to2$ processes started to be calculated,
including photon+jet~\cite{gammaJetAtNLO}, heavy-quark pair
production~\cite{HQatNLO} and dijet production~\cite{Aversa:1988vb,Giele:1993dj}.
$2\to3$ processes started to follow ten years later, notably $Wb\bar
b$ production~\cite{WbbNLO}.
Extrapolating, it should come as no surprise that it was around 2009
that $2\to 4$ processes began to appear, with two calculations for
W+3\,jets~\cite{W3jNLO} and three for the production of two heavy
$q\bar q$ pairs~\cite{ttbbNLO}.

\begin{figure}
  \centering
  \begin{minipage}{0.48\linewidth}
    \includegraphics[width=\textwidth]{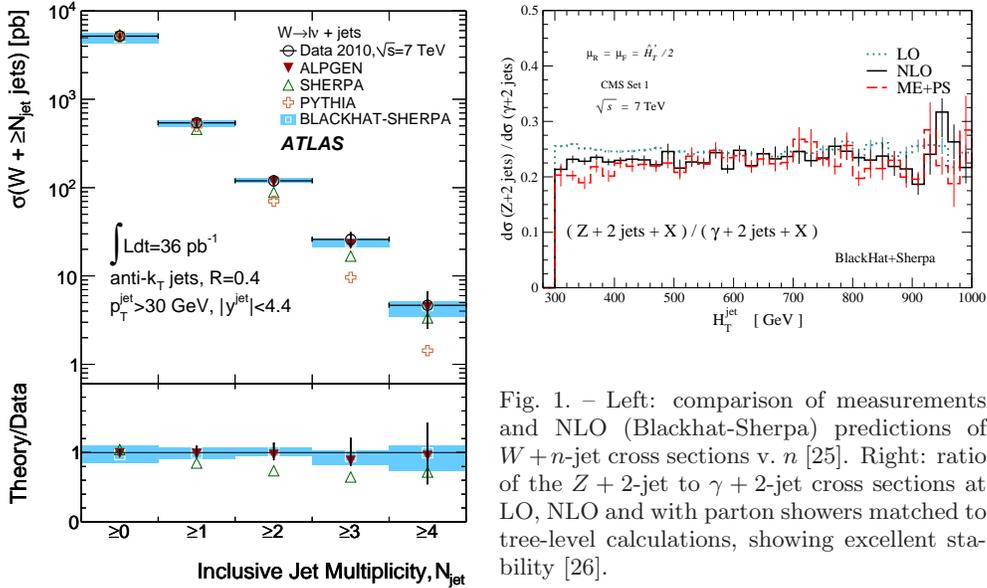}\hfill
  \end{minipage}
  \begin{minipage}{0.48\linewidth}
    \includegraphics[width=\textwidth]{1106-1423-nlo-z-gamma-ratio-v-ht.eps}\\
  \caption{Left: comparison of measurements and NLO (Blackhat-Sherpa) predictions of
    $W+n$-jet cross sections v.\ $n$~\cite{Aad:2012en}.
    Right: ratio of the $Z+2$-jet to $\gamma+2$-jet cross sections
    at LO, NLO and with parton showers matched to tree-level
    calculations, showing excellent stability~\cite{Bern:2011pa}. 
  }
  \end{minipage}
  \label{fig:W4j}
\end{figure}

One might deduce that one should then wait until around 2019 for the
background process that we mentioned above, $Z$+4~jets, a $2\to5$
process.
Yet, remarkably, its calculation appeared 8 years ahead of schedule,
in 2011~\cite{Z4jNLO} (another $2\to5$ process, W+4~jets, came out earlier, in
2010~\cite{W4jNLO}, and is in excellent agreement with data,
cf.~Fig.~\ref{fig:W4j} left).
There have even been first calculations of processes with a complexity
equivalent to that of a $2\to6$ process~\cite{ee7j}.
While it is beyond the scope of this review to describe the many
ingenious innovations behind this progress, it should be said that a
number have benefited from recent progress in understanding how to
fully carry out a 20-year old dream~\cite{Bern:1994zx} of sewing
together tree amplitudes to obtain loop graphs (for more details see
\cite{Ellis:2011cr}).

Given this progress, it is natural to ask if it is being used by the
experiments.
At first sight, when opening one of the many SUSY search papers from
ATLAS or CMS one is initially disappointed: nearly all the results
rely either on matrix-element (tree-level) calculations supplemented
with parton showers (ME+PS) or ``data-driven'' estimates of
backgrounds.
Part of the reason is that pure NLO calculations provide information
about partons, whereas experiments can only sensibly input hadrons to
their detector simulations.
Yet, digging deeper, there are cases where cutting-edge NLO results
are already having an impact. 
An example is~\cite{CMS-PAS-SUS-1104}, which compares data both to
ME+PS and to data-driven background estimates, relying on the latter
for its final SUSY exclusion limits.
``Data-driven'' sounds as if it is altogether independent of
theorists. 
In this specific case, for estimating the $Z$+jets background the idea
is to measure the $\gamma$+jets cross section (instead of a direct
measurements of $Z$'s, which suffers from the low $Z\to \ell^+\ell^-$
branching ratio) and then to use NLO predictions for the ratio of
$\gamma$+jets to $Z$+jets to deduce the expected measured $Z$+jets
background. 
Many experimental systematics such as jet-energy scale are common to
both and therefore cancel in the ratio; meanwhile the theoretical
prediction is extremely stable, Fig.~\ref{fig:W4j} (right). 
So, the data-driven method here is actually a clever way of exploiting
precisely known aspects both theory and experiment, while minimising
the impact of their intrinsic limitations.
More generally, data-driven methods don't always (or even often) use
NLO, but they do quite often involve this idea of finding a way to
combine the best of theory and experiment.

\section{Systematically matching showers and NLO}

Despite the power of data-driven methods, there remain many cases
where the experiments do need a direct, quality prediction of
hadron-collider processes.
This is crucial in many Higgs searches, which nearly always rely on
precise hadron-level predictions of the signal, and also often of the
backgrounds.
And it was the case also for the analysis that led to the $W$+2-jet anomaly reported by
CDF~\cite{Aaltonen:2011mk}, but not found by D0~\cite{Abazov:2011af}.
One of the standards for collider predictions involves the matching of
tree-level matrix-element calculations with parton showers and it is
to such predictions, passed through detector simulations, that the 
CDF and D0 W+2j results were compared. 

Combining tree-level (i.e.\ LO) calculations and parton showers is
relatively easy nowadays thanks to automated tools for tree-level
predictions of essentially any standard-model process (e.g.\
MadGraph~\cite{Alwall:2007st}, Alpgen~\cite{Mangano:2002ea},
Sherpa~\cite{Gleisberg:2008ta}) and methods such as
MLM~\cite{Alwall:2007fs} and CKKW~\cite{Catani:2001cc} 
matching, which address the issue of combining tree-level calculations
for different multiplicities, while avoiding the double counting that
would be caused by the fact that parton-showers themselves generate
extra emissions (for a recent review see~\cite{Buckley:2011ms}).

A concern with any calculation based on tree-level methods is that it
is essentially a leading-order calculation, yet we know that NLO
corrections can be large.
NLO calculations can also be combined with parton showers, through the
MC@NLO~\cite{Frixione:2002ik} or POWHEG~\cite{Nason:2004rx} methods
(see also~\cite{Nason:2012pr}),
generally with a significant improvement in precision; this is less
straightforward than for tree-level calculations, because of the need
for 1-loop matrix-elements and because of extra issues of double
counting that arise at NLO.
Until recently it had always been done on a case-by-case basis,
usually limited to cases with low multiplicities (e.g.\ just W
production, with no jets other than that from the NLO radiation).
A major development of the past couple of years has been that a huge
number of extra processes has become available. 
Two approaches have been taken: one is the POWHEG-Box~\cite{Alioli:2010xd},
which provides well-documented infrastructure for taking existing NLO
calculations and turning them into a POWHEG type calculation (a
related approach been taken in Sherpa~\cite{Hoche:2010pf}); the
other is the aMC@NLO~\cite{Hirschi:2011pa} approach, which builds on
MadGraph so as to automatically calculate NLO (loop and tree-level)
matrix elements and combine them with parton showers with the MC@NLO
method.

\begin{figure}
  \centering
  \begin{minipage}[t]{0.57\linewidth}
    \phantom{x}\mbox{ }\\[-1em]
    \includegraphics[width=\textwidth]{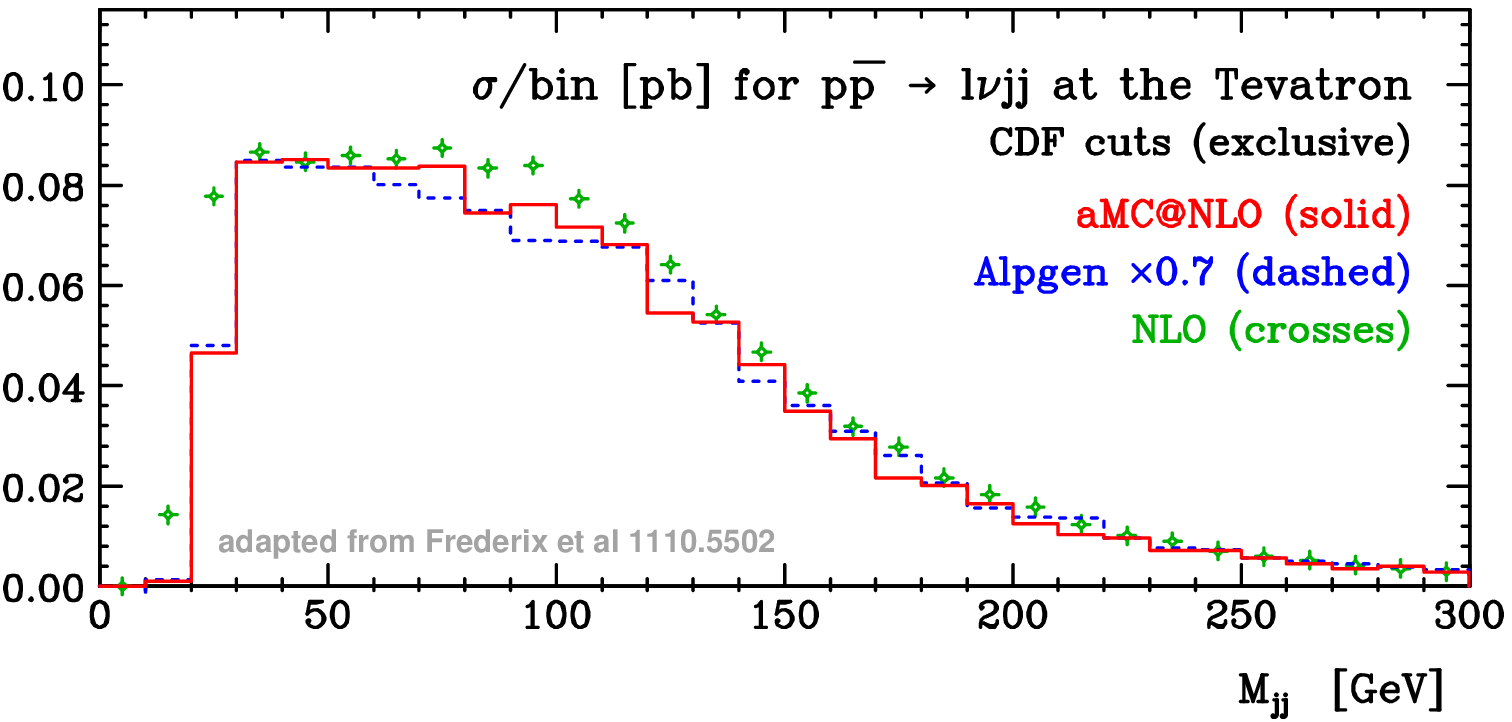}\\
    \mbox{}\hfill\includegraphics[width=0.99\textwidth,height=0.35\textwidth]{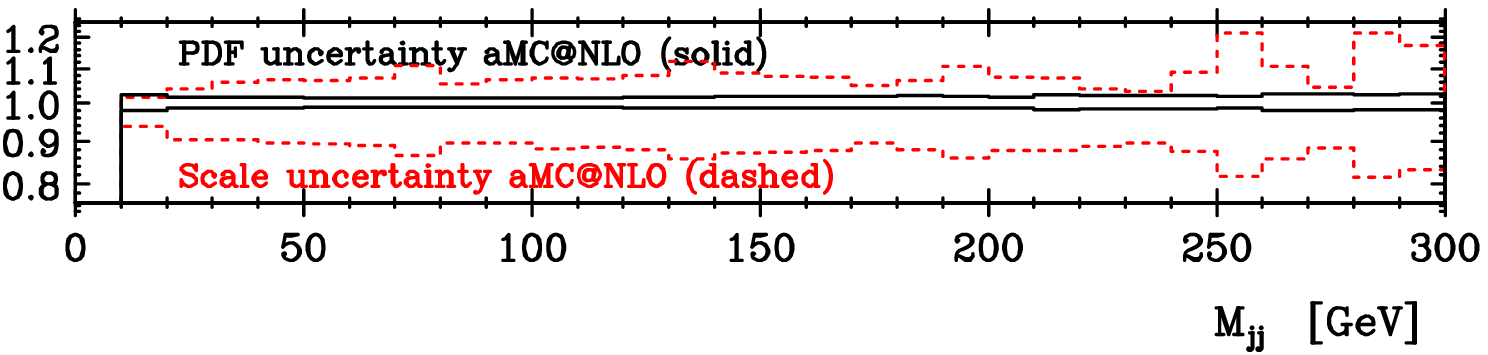}
  \end{minipage}
  \begin{minipage}[t]{0.4\linewidth}
    \phantom{x}\mbox{ }\\[-1em]
    \includegraphics[height=\textwidth]{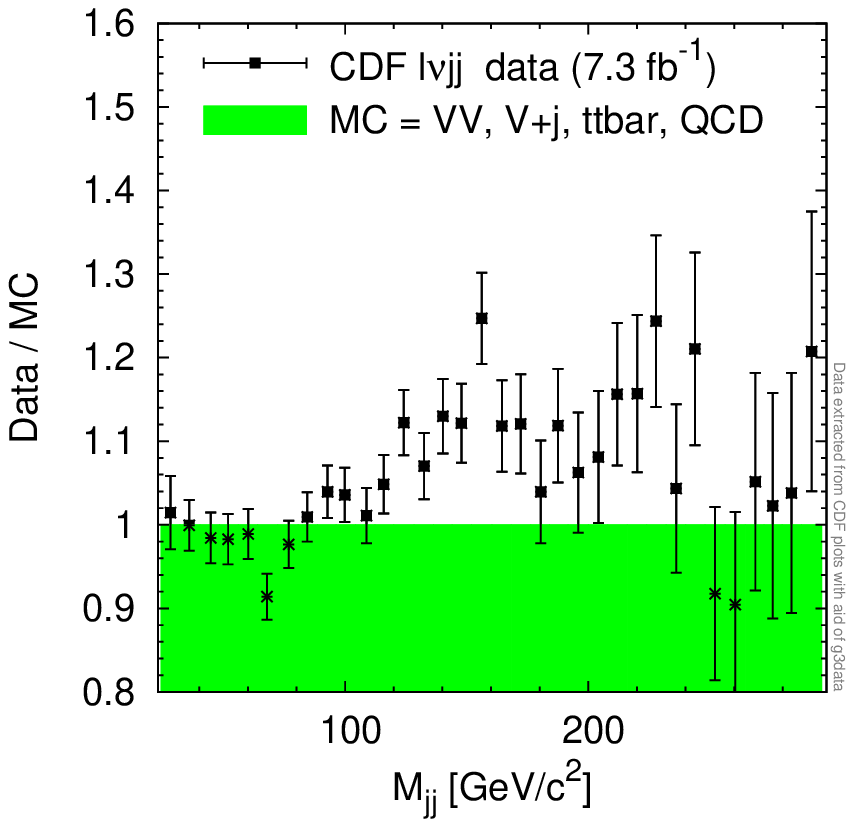}
  \end{minipage}
  \caption{Left: aMC@NLO calculation of the dijet mass spectrum in
    W+2j events, compared to Alpgen (ME+PS) and pure NLO; also shows
    are the PDF and scale uncertainties. Right: the CDF data as read
    from the plots of \cite{Aaltonen:2011mk}, shown as a ratio to the expected
    background (also read from the plots), instead of the usual
    difference between data and background.}
  \label{fig:W2j}
\end{figure}

A calculation of the W+2j process with the CDF cuts has been one of
the headline applications of the aMC@NLO method, with results shown in
Fig.~\ref{fig:W2j} (left).\footnote{Showered NLO W+3-jet results have
  since then become available with the MC@NLO method in
  Sherpa~\cite{Hoeche:2012ft}. }
The aMC@NLO and Alpgen results (the latter after a rescaling to
account for an overall NLO K-factor) agree remarkably well in this
case, suggesting that predictions are relatively stable.
Interestingly, aMC@NLO and Alpgen bear more similarity to each other
than either does to the pure NLO results, highlighting the importance
of showering effects.

One advantage of NLO calculations (whether with a shower or not) is
that scale variations can offer a reasonable way of estimating
uncertainties. 
These uncertainties are shown also in Fig.~\ref{fig:W2j} (left) and
are of the order of $10-15\%$.
For comparison, the right-hand plot of Fig.~\ref{fig:W2j} shows the
CDF data as a {\it ratio} to the expected background. What appears as
a peak around $150\GeV$ in the usual plots
of (data$\,-\,$background), here appears as the start of a broad $10\%$
excess starting around $140\GeV$, with only the barest hints of a
peak around $150\GeV$. 
It is not my intention to claim that background uncertainties
are responsible for CDF's anomaly, but simply to provide a reminder
that when looking at effects of the order of
$10\%$, we enter a region where our ability to predict backgrounds
sufficiently accurately can start to become a serious issue.
Of course, a number of explanations have been proposed to
explain the anomaly (beyond the scope of this talk). 
Ultimately, however, it is almost certainly impossible for outsiders
to resolve this issue, and one can only hope that the CDF and D0
experiments will at some point have the means to bring closure to the
question.

\section{Going beyond the limitations of NLO}

There are at least two avenues to go beyond the limitations of NLO
calculations.
On one hand one can aim for high precision, for example with NNLO.
On the other, in the context of searches, one can try to make signals
emerge more clearly.

\subsection{Next-to-next-to-leading order}

NNLO is available for all processes involving the production of a
single vector or Higgs boson, but until recently was mostly inexistent
for processes with two bosons or with coloured particles in the final
state.
It is crucial not just in prediction backgrounds, but also in deriving
information from ``signals'' --- for example for extracting parton
distribution functions from $W$ and $Z$ production differential cross
sections; and, once (if) the Higgs-boson is discovered, for
determining its couplings to the rest of the standard model.

\begin{figure}
  \centering
  \includegraphics[width=0.48\textwidth,height=0.42\textwidth]{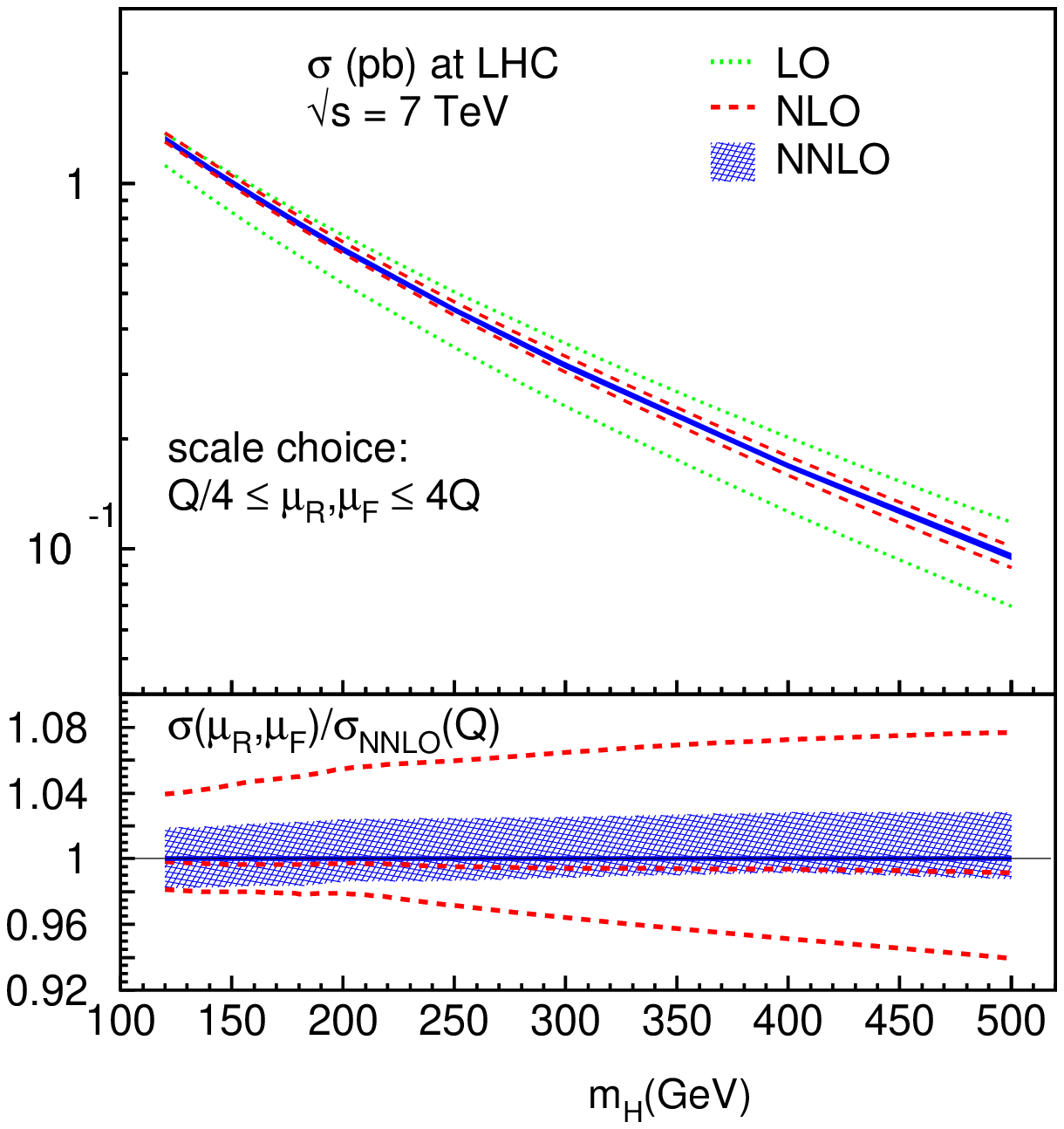}\hfill
  \includegraphics[width=0.48\textwidth,height=0.42\textwidth]{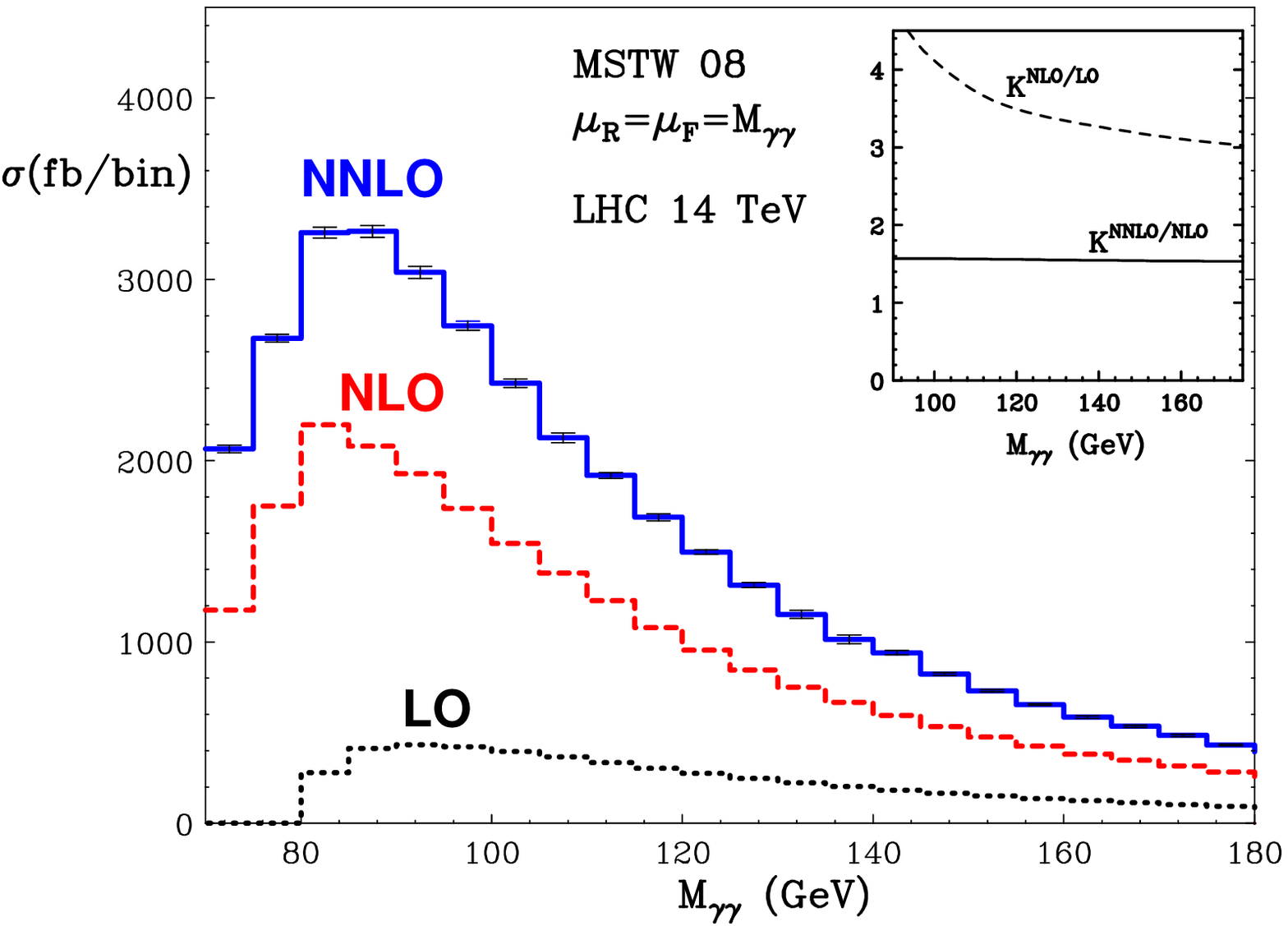}\\
  \caption{Recent NNLO predictions for inclusive vector-boson fusion~\cite{Bolzoni:2010xr}
    (left) and differential $\gamma
    \gamma$ production~\cite{Catani:2011qz} (right).} 
  \label{fig:NNLO}
\end{figure}

The past couple of years has seen several new NNLO calculations and
they can roughly be divided into two classes: one class is that where NNLO
delivers its promise of high precision, for example the inclusive
calculation of vector-boson fusion
Higgs-production~\cite{Bolzoni:2010xr} or the differential
calculation of WH production~\cite{Ferrera:2011bk}. The
former is shown in Fig.~\ref{fig:NNLO} and, for once, the convergence
is just as one would expect with a coupling $\sim 0.1$.

A second class of processes has more sobering results: Fig.~\ref{fig:NNLO} (right) shows
the recent NNLO $\gamma \gamma$~\cite{Catani:2011qz} prediction where
one sees $50\%$ corrections in going from NLO to NNLO.
Yet another case where NNLO appears to face difficulties and that has
seen extensive discussion is for the prediction of the efficiency of
jet vetoes in Higgs production~\cite{Dittmaier:2012vm}.

Various physics mechanisms can lead to this poor behaviour of the
perturbative series: Sudakov logarithms when one puts excessively
stringent cuts on the final state; threshold logarithms when one is
limited by the available partonic centre of mass energy; and the
appearance of partonic scattering channels at NLO or NNLO that weren't
possible at lower order (e.g. $qq \to qq \gamma \gamma$).
Still despite progress on individual aspects (e.g. Higgs Sudakov
boson~\cite{Bozzi:2005wk,Cao:2009md} and
jet~\cite{Banfi:2012jm,Becher:2012qa,Tackmann:2012bt,Banfi:2012yh} $p_t$ 
resummations, Higgs 
threshold
resummations~\cite{Catani:2003zt,Kidonakis:2007ww,Ahrens:2008qu}, or
the combination of 
processes with different multiplicities~\cite{Rubin:2010xp}), one cannot
help but wonder whether with more insight we might not obtain
predictions that just systematically converge better.

I cannot close this section without mentioning a result that appeared
only after my talk, the NNLO corrections to $q\bar q \to t\bar
t$~\cite{Baernreuther:2012ws}. It's the first time that a NNLO
calculation appears for a process involving coloured particles in both
initial and final states and as such is a groundbreaking achievement.
It brings significantly improved precision, at the $3\%$ level
(roughly a factor of $3$ improvement over NLO), to the prediction of
the Tevatron $t\bar t$ cross section.%
\footnote{%
  It is also interesting because there were a number of predictions for
  the $t\bar t$ cross section based on threshold resummation, even
  though Tevatron $t\bar t$ production is not strictly at threshold.
  A verification of how well these performed can provide useful insight
  into the applicability of different approaches to threshold
  resummation, possibly also in other not-really-threshold contexts such
  as LHC Higgs production.}

\subsection{Rethinking searches}

It can be tempting to say that if only we knew collider backgrounds
better we would significantly improve the reach of a range of
searches. 
But I believe it's just as important to ask the question of whether
using our understanding of QCD can help us to find better analysis
techniques, techniques that improve signal-to-background ($S/B$)
ratios (and sometimes also $S/\sqrt{B}$) and so reduce the impact of
background uncertainties.

One example of this is in the search for $b\bar b$ decays of a light
Higgs boson when produced in association with a W or Z.
Long thought to be inaccessible due to huge backgrounds, this channel
became possible again after the results of \cite{Butterworth:2008iy},
which showed how to improve signal to background ratios by going to
high $p_t$ (backgrounds drop faster than signal) and developing
appropriate ``boosted-Higgs'' jet-substructure reconstruction methods.
Since then, the idea of going to high $p_t$ has been adopted
in~\cite{ATLAS-VHbb,CMS-VHbb}, though there are insufficient data so
far to truly benefit from the new jet substructure methods.
In parallel, this and other early work has spurred a whole new field
of investigation into event-topologies with boosted tops, vector and
Higgs bosons, and BSM particles, together with extensive further
developments on the jet-substructure analysis techniques.
Several reviews cover this progress in
detail~\cite{Abdesselam:2010pt,Altheimer:2012mn,Plehn:2011tg}.

While ``boosted-object'' searches have garnered much of the attention
recently, progress is also being made in better signal detection in
other areas too: quark/gluon discrimination has recently seen a
renewal of interest~\cite{Gallicchio:2011xq} and may have applications
in many searches; and there is clearly scope for novel methods even in
``standard'' SUSY searches~\cite{Hook:2012fd}.

A general question that remains open here is whether such improvements
will continue to come mainly from an intuitive understanding of QCD
v.\ BSM differences, as has largely been the case so far, or whether
there is a benefit to be had also from more quantitative, analytical
insight into how signals and backgrounds differ.

\section{Conclusions}

\begin{figure}
  \centering
  \includegraphics[width=0.93\textwidth]{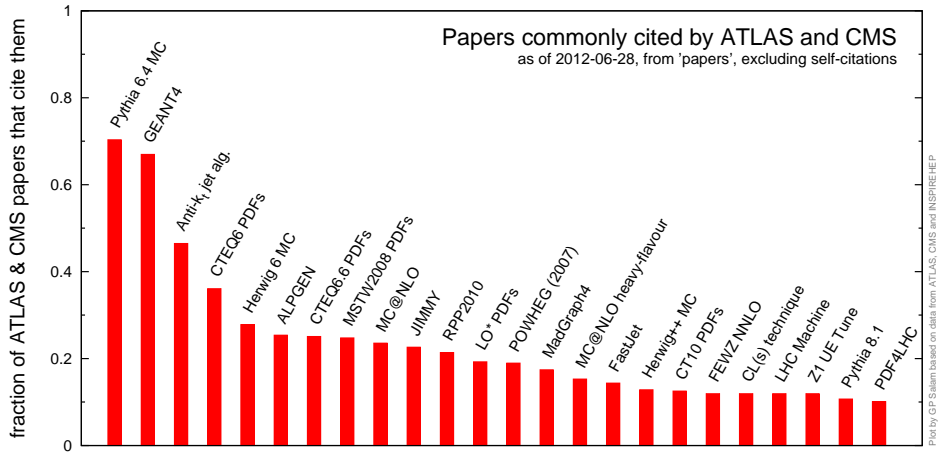}
  \caption{The papers most frequently cited by the  ATLAS and CMS
    collaborations (excluding citations to ATLAS and CMS), showing the
    fraction of ATLAS and CMS papers that refer to each one. The lists
    of ATLAS and CMS papers were taken from their respective websites and
    the citations data from INSPIRE.}
  \label{fig:spires}
\end{figure}

Many of the cutting-edge QCD research results discussed here have come
with an initial focus on one or other specific class of search or
application.
It is interesting to look also at what the LHC experiments use across
the board, on a day-to-day basis.
Fig.~\ref{fig:spires} shows the papers most commonly referred to by
ATLAS and CMS and, for each, the fraction of the collaborations'
articles that refer to them.
Some aspects of this graph are unsurprising, such as the overwhelming
role played by Pythia.
Other aspects provide a reminder that such citations data should be
interpreted with an abundance of caution: one can't help but notice
the position of the ``LHC Machine'' relative to Pythia.\footnote{Of
  course, Pythia is easier to run.}
Still, it is striking that of the 24 articles shown (those cited by
more than $10\%$ of the collaborations' papers), 20 stem from the QCD
community, a tribute to the key role being played by QCD at the LHC.
%

\acknowledgments

I am grateful to Viviana Cavaliere, Jan Winter and Wenhan Zhu for
numerous discussions related to the CDF W+dijet anomaly and to Matteo
Cacciari for comments on the manuscript.
I also wish to thank the organisers for the stimulating environment
provided during the workshop and both the organisers and the French
Agence Nationale de la Recherche (grant ANR-BLAN-2009-060) for
financial support.

\end{document}